\documentclass[%
 aip,
 apl,
 amsmath,amssymb,
 reprint,%
]{revtex4-1}
\usepackage{mathptmx,bm}

\usepackage{graphicx}% Include figure files
\usepackage{dcolumn}% Align table columns on decimal point
\usepackage[utf8]{inputenc}
\usepackage[T1]{fontenc}

\begin{document}

\preprint{AIP/123-QED}

\title[Sample title]{Observation of Kerr nonlinearity and Kerr-like nonlinearity induced by terahertz generation in LiNbO$_3$ 
}
% Force line breaks with \\

\author{Bo Wang}
 \affiliation {Beijing National Laboratory for Condensed Matter Physics, Institute of Physics, Chinese Academy of Sciences, Beijing 100190, China}%Lines break automatically or can be forced with \\
  \affiliation {School of Physical Sciences, University of Chinese Academy of Sciences, Beijing 100049, China
}
\author{Chun Wang}%

 \affiliation {Beijing National Laboratory for Condensed Matter Physics, Institute of Physics, Chinese Academy of Sciences, Beijing 100190, China}%Lines break automatically or can be forced with \\
  \affiliation {School of Physical Sciences, University of Chinese Academy of Sciences, Beijing 100049, China
  }

\author{Li Wang}
\affiliation {Beijing National Laboratory for Condensed Matter Physics, Institute of Physics, Chinese Academy of Sciences, Beijing 100190, China}%Li
\author{Xiaojun Wu}

 \email{xiaojunwu@buaa.edu.cn}
 \affiliation{{Beijing Key Laboratory for Microwave Sensing and Security Applications, School of Electronic and Information Engineering, Beihang University, Beijing, 100083, China}
}%

\date{\today}% It is always \today, today,

\begin{abstract}

Femtosecond laser pulses interacting with LiNbO$_3$ has been widely applied in intense terahertz generation, excitation and imaging of phonon polaritons and modulating light. It is of great importance to study the nonlinearities induced by femtosecond laser pulses in LiNbO$_3$. Here, we demonstrate that both Kerr nonlinearity and Kerr-like nonlinearity induced by terahertz wave occur under the interaction between laser pulses and LiNbO$_3$ using optical pump-probe techniques. We show that the contribution of Kerr-like nonlinearity to pump-probe spectra varies with the interaction length because of the phase mismatching between terahertz wave and laser pulse. We also observed the excitation of the low frequency phonon polaritons. Experimental results are in consistent with theoretical calculation. 
\end{abstract}

\maketitle

Since the invention of laser, nonlinear optics has always been drawing attentions of researchers. LiNbO$_3$ is one of the most important nonlinear optical crystals \cite{David2005} with applications in waveguides\cite{Alibart2016} and high speed modulators \cite{Wooten2000}.  Especially in terahertz (THz) science and technology , LiNbO$_3$  is widely used in intense THz sources \cite{Hafez2016, Hebling2004} under the illumination of femtosecond laser pulses with the tilted pulse front technique. Femtosecond laser pulses induced nonlinearity in LiNbO$_3$  is of great significance to study.

In recent decades, one of  interests in LiNbO$_3$  is excitation and detection of phonon polaritons with pump-probe techniques in which one or two femtosecond laser beams excite phonon polaritons and a weaker probe beam detects the refractive index changes caused by them \cite{Bakker1998}. However, the induction of coherent phonon polaritons is not the only nonlinear optical effect when femtosecond laser pulses interact with electro-optic crystals. Third order optical nonlinearity exists in media with any symmetry \cite{Shen1984}. Kerr effect is one of the third order nonlinear optical effects with an induced nonlinear index  which makes the refractive index to be defined as $n=n_0 +n_2 I$, where $I$ is the intensity of laser beam.  Besides Kerr effect, a cascaded second-order nonlinear optical effect called Kerr-like nonlinearity \cite{Bosshard1995} occurring in non-centrosymmetric crystals will induce a similar $n_2$. Caumes et al \cite{Caumes2002}  have reported a THz waves related Kerr-like effect in ZnTe where femtosecond laser pulses generate THz waves via optical rectification and the generated THz waves modify the refractive index at optical frequency via electro-optic effect. A combination of the optical rectification and electro-optic effect induces an equivalent $n_2$.  It is necessary to distinguish Kerr and Kerr-like effect in LiNbO$_3$ because this kind of crystals has important applications in THz devices.

We present an optical pump-probe experiment on LiNbO$_3$. Unlike previous studies where only oscillations from phonon polaritons were investigated, we focus on the  signals near zero-time delay. We demonstrate that both Kerr and Kerr-like effects contribute to this signal. Although a large velocity mismatch exists between laser pulses and THz waves in LiNbO$_3$, the obtained spectra still have obvious Kerr-like  properties, which is inconsistent with previous  results\cite{Caumes2002}  where only THz waves almost fulfilling the phase matching condition contribute obviously to $n_2$. 

\begin{figure}
\includegraphics{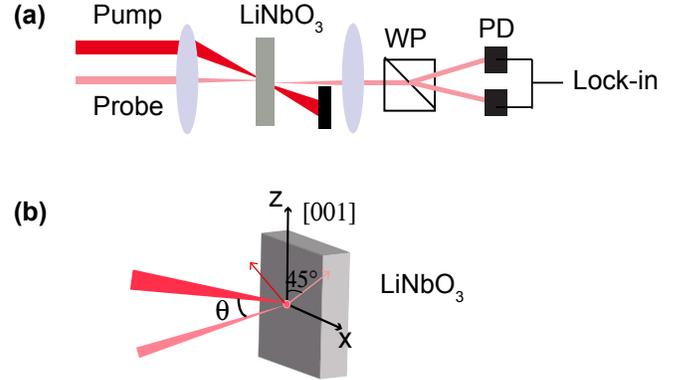}% Here is how to import EPS art
\caption{\label{fig:1}(a) Pump probe experimental setup. (b) Geometries of sample and polarizations of the pump and probe beams.}
\end{figure}

Fig. 1 shows the setup for optical pump-probe experiment.  A congruent MgO: LiNbO$_3$ (MgO 6\%) wafer was used in our experiment. LiNbO$_3$ belongs to uniaxial crystals.  Our sample was cut with its optic axis( defined as  z axis) in the plane of surface (x-z plane) and its thickness of 2 mm. Both sides of our sample were polished.   Femtosecond laser pulses generated by a commercial Ti:Sapphire oscillator  with  duration of 70 fs, repetition frequency of 80 MHz, central wavelength of 800 nm, were split into a pump beam and a probe beam. The pump beam, after being delayed by an optical stage  and modulated by an optical chopper, was focused onto the surface of the LiNbO$_3$ wafer. The probe beam was focused at the same place as the pump beam, with the polarization perpendicular to that of the pump beam to decrease the coherent spike signal \cite{Luo:09}. The polarization of the probe beam was 45$^\circ$ off the optical axis, and the polarization variation  induced by the pump beam was detected by a combination of a quarter wave-plate, a Wollaston prism and a balanced detector. The signal as a function of delay time was collected by a phase sensitive lock-in amplification  technique. The pump beam was non-collinear with the probe beam. The average power of the pump beam was $\thicksim$   435 mW and that of the probe beam was $\thicksim$ 20 mW. The spot size of the pump beam was larger than that of the probe beam. Before scanning the optical stage line, we blocked the pump beam and adjusted the quarter wave plate to make the output signal to be zero.  We  conducted all the experiments  at the room temperature.

\begin{figure}

\includegraphics{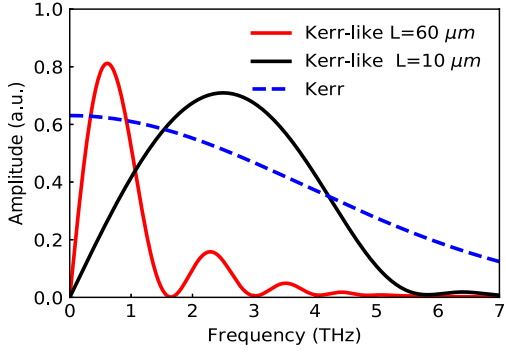}% Here is how to import EPS art
\caption{\label{fig:2} Calculated pump probe spectra originated from Kerr (blue dash line) effect, Kerr-like effect when L=60 $\mu m$  (red line) and Kerr-like effect when L=10 $\mu m$(black line).}

\end{figure}

Firstly, we demonstrate theoretically that both Kerr and Kerr-like nonlinearities contribute to the pump-probe spectra.  LiNbO$_3$ crystals at the room temperature belong to 3m point group \cite{Weis1985}  and their third order nonlinear susceptibility $\bm{\chi^{(3)}}$ has 14 independent nonzero elements and,  their second order nonlinear susceptibility $\bm{\chi^{(2)}}$ has 4 independent  nonzero elements \cite{Boyd2008}. In our experiment, the pump beam induces a  variation of dielectric tensor via $\bm{\chi^{(3)}}$ given by

\begin{equation}
\Delta  \bm{\varepsilon} ^{Kerr} \propto \frac{1}{2} \left[\begin{array}{cc} \chi_{xxxx}+\chi_{xxzz}& 2\chi_{xzxz} \\2\chi_{xzxz} & \chi_{zzxx}+\chi_{zzzz}\end{array}\right]E^2 ,
\end{equation}
 where $\chi_{ijml}$ $(i,j,m,l=x,z)$ is the component of $\bm{\chi}^{(3)}$ and $E$ is the electric field of the laser pulse. The induced $\Delta \bm{\varepsilon} ^{Kerr}$  modifies  the polarization of the probe beam. Since the $ \bm{\chi}^{(3)}$  effect here is instantaneous, for each time delay $\tau$  , the lock-in gives rise to  a signal\cite{Caumes2002, Tian2008}:  $S^{Kerr}(\tau) \propto\int I^{pu}(t)I^{pr}(t-\tau)dt$ , which is the  correlation of the pump and probe beams, and has no negative part.  The Fourier transform of  $S^{Kerr}(\Omega)$ is shown in Fig. 2 (blue dash line).  It  is a descending function with a finite dc component.

The second order nonlinear susceptibility $\bm{\chi}^{(2)}$ of LiNbO$_3$ contributes to the generation of THz electric fields via optical rectification of laser pulses,  and the generated THz waves induce a modification of the dielectric tensor via electro-optic effect. The sum effect gives rise to the Kerr-like effect and the induced change of the dielectric tensor is given by 

\begin{equation}
\Delta \bm{\bm{\varepsilon}} ^{Kerr-like} \propto \frac{1}{2} \left[\begin{array}{cc} \chi_{zxx}\chi_{zxx}& 2\chi_{xzx} \chi_{xxz}\\2\chi_{xzx}\chi_{xzx} & \chi_{zzz}\chi_{zzz}\end{array}\right]E^2,
\end{equation}
where $\chi_{ijm}$ $(i,j,m=x,z)$ is the component of $\bm{\chi^{(2)}}$. Here, the relations $\chi_{mij}= \chi_{mji}$ and $r_{ijm} \propto \chi_{mij}$ are invovled, where $ r_{ijm} $ is the electro-optic coefficient. Eq. (2) means that   $\Delta \varepsilon ^{Kerr-like}_{ij} $ can be induced with THz electric field $E_m^{THz}$ via  $r_{ijm}$, where the $E_m^{THz}$ is generated by femtosecond laser pulses via $\chi_{mij}$.  The $\Delta \bm{\varepsilon} ^{Kerr-like} $  has a similar form with the $\Delta \bm{ \varepsilon} ^{Kerr} $ and will also modify the polarization of the probe beam. 

The generated THz wave has a propagation effect in LiNbO$_3$. It is well known that there exists a large velocity mismatch between the laser pulses and THz waves in LiNbO$_3$. The velocity mismatch will greatly influence both the generation and probe processes, which makes the spectrum shape of Kerr-like effect different from that of the Kerr effect. The spectrum due to the  generation and detection of the THz waves is calculated according to  Ref. \onlinecite{Schneider:06}. For the generation process, the THz electric field is expressed as

\begin{equation}
E(\Omega,z)=\frac{\mu_0 \Omega \chi^{(2)}_{eff} I(\Omega)}{i n(\omega_o) [n(\Omega)+n_g]} \frac{e^{ik_0n(\Omega)z}-e^{ik_0n_g z}}{k_0[n(\Omega)-n_g]},
\end{equation}
where $\chi^{(2)}_{eff}$ is the effective second order coefficient which is assumed as a constant;  $n(\omega_o)$ and $n(\Omega)$ are the refractive index at optical frequency and THz frequency range, respectively; $n_g$ is the group refractive index of laser pulses;  z is position; $I(\Omega)$ is the spectrum of laser envelope; $k_0=\Omega/c$, where $c $ is the light velocity in vacuum.  For the probe process, the spectrum of the signal obtained from the lock-in is described as 
\begin{equation}
S^{Kerr-like}(\Omega,L)=k_0 \chi^{(2)}_{eff} I(\Omega) \frac{1-e^{ik_0[n(\Omega)-n_g]L}}{-i[n(\Omega)-n_g]}E(\Omega,L).
\end{equation}
In the  Ref. \onlinecite{Schneider:06},  $L$ in Eq. (4) denotes the sample thickness. In our case, since the pump beam is non-collinear with the probe beam, L is defined as the interaction length, which can be controlled by the angle  $\theta$ between the pump and probe beam as shown in Fig. 1. 

To investigate the properties of the Kerr-like nonlinearity in the pump-probe experiment, we calculated the spectra when $L=60$  $\mu m$ and $L=10$  $\mu m$ according to Eq. (4). The parameters $n(\Omega)$ and $n(\omega_o)$  are given in Ref. \onlinecite{Zinov2007}. The calculated spectra  $S^{Kerr-like}(\Omega)$ due to Kerr-like effect are also illustrated in Fig. 2.  There exist two significant differences between spectra from the Kerr effect and the Kerr-like effect. First, $S^{Kerr}(0)$ has a finite value while  $S^{Kerr-like}(0)$ vanishes. Second, $S^{Kerr-like}(\Omega)$ has an interference structure which is originated from the propagation of laser pulse and THz waves. The interference structure is dependent on the interaction length $ L$. The smaller $L$ makes a smaller structure. The whole spectrum $S(\Omega)$ is the sum of $S^{Kerr-like}(\Omega)$ and $S^{Kerr}(\Omega)$ . 

\begin{figure*}
\includegraphics[width=17 cm, height=5.5 cm]{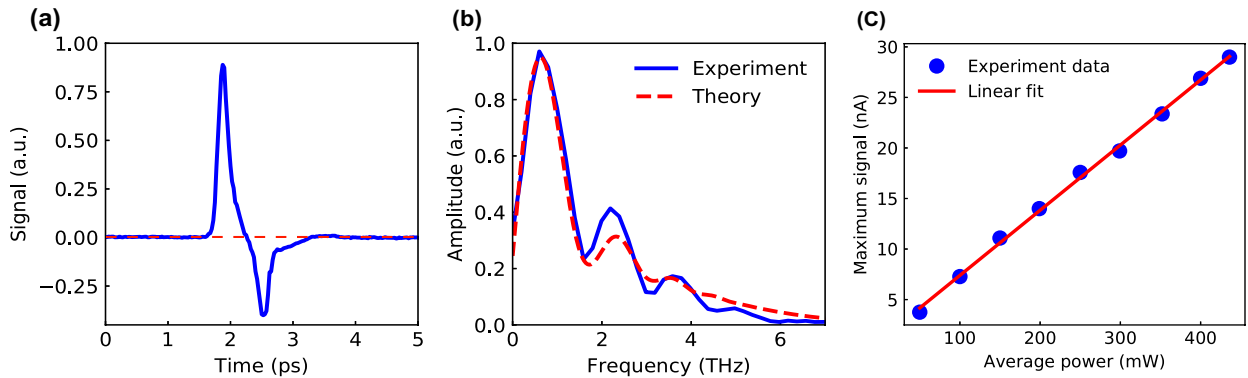}% Here is how to import EPS art
\caption{\label{fig:3}  (a)  A typical waveform obtained from the pump-probe measurement. The red dash line is the guide line of zero signal.  (b) The corresponding Fourier transform spectrum (blue line) and the calculated spectrum (red dash 
line). (c) Peak value of the waveform as a function of the pump power (blue circles) and the linear fit (red line) for the experimental data.}

\end{figure*}

\begin{figure*}

\includegraphics[width=17 cm,height=5.5 cm]{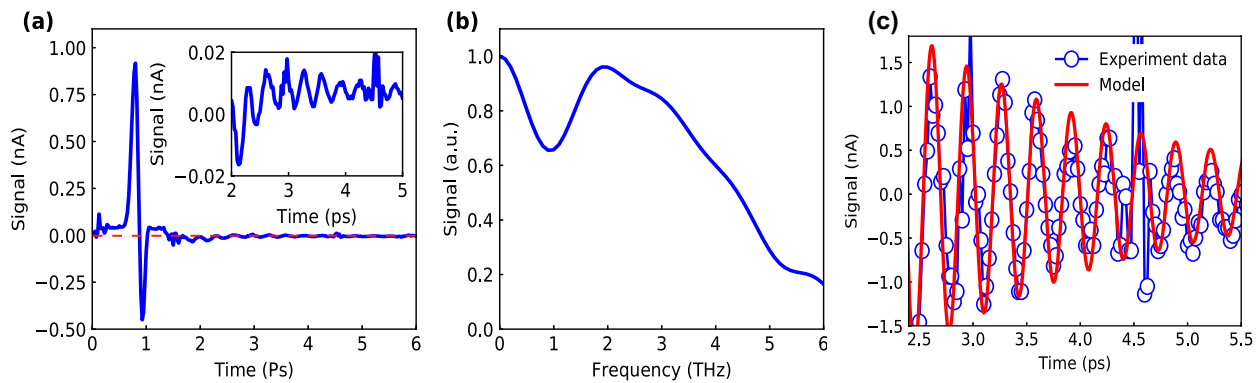}% Here is how to import EPS art
\caption{\label{fig:4} (a)  A pump-probe signal with a longer tail, and (b) its corresponding spectrum when the angle between the pump and the probe beams is  $\thicksim$ 10$^\circ$. Inset: the zoom in view of the oscillation waveform. (c) A damped oscillator model is employed to well fit the oscillatory waveform in 2-5 ps.}

\end{figure*}

To evidence our  theoretical analysis, we conducted pump-probe experiment at two different $\theta$. A typical experimental signal recorded when $\theta  \approx 5 ^\circ$  is depicted in Fig. 3(a). It has a one-cycle pulse  with duration about 1.5 ps which has a negative part, indicating that THz related Kerr-like effect contributes to the signal. Fig. 3(b) shows the Fourier Transform spectrum of the waveform. This spectrum spans up to 6 THz and has an interference structure which is similar to that in Fig. 2.

According to our calculation, the interference structure originates from the large velocity mismatching between  the THz waves and  laser pulses. Besides, the spectrum has a nonzero dc component. These results indicate that both Kerr and Kerr-like nonlinearity play roles in $n_2$. By adjusting the interaction length and the ratio between Kerr and Kerr-like spectra, we can reproduce the measured spectrum, as is shown in the red dash line in Fig. 3(b).  We also conducted power dependent experiments. Fig. 3(c) shows the peak value of the waveforms as a function of the laser  average power of pump beam. The line represents the linear fit of the experimental data. It is obvious that the pump-probe signal scales linearly with the intensity, which is expected by both Kerr and Kerr-like nonlinearities.

 We increased the angle between the pump beam and the probe beam to $\thicksim$  $ 10^\circ $ to decrease the interaction length. Figure 4(a) shows the waveform when the interaction length is decreased. There is a big one-cycle pulse followed by a small oscillation signal as shown in the inset. The duration of the main pulse is $\thicksim$ 1 ps, which is a little smaller than that in Fig. 3(a).
The  Fourier transform spectrum  of the main pulse is plotted in Fig. 4(b). It spans more than 6 THz,  and the interference structure is not obvious. Besides, this spectrum has larger high frequency components than that shown in Fig .3(b). These results are consistent with our calculation for the smaller interaction length.   

Following the main pulse in Fig. 4(a), there is a long oscillation behavior. This is due to the coherent excitation of phonon polaritons.  Pioneering works about coherent phonon polaritons excited by femtosecond laser pulses via impulsive Raman scattering  have been reported by Bakker and Nelson el al in Ref. \onlinecite{Bakker1998,Dougherty1992}. The oscillation  is perfectly fitted by a damped harmonic oscillator function, seen in Fig. 4(c). The frequency of the phonon polaritons, obtained from the fitting of the experimental data, is $\thicksim$ 3 THz, which is consistent with previous report\cite{Blake2017}.

Even on a bad phase matching condition, we can still  observe THz related Kerr-like nonlinearity when femtosecond laser pulses interacting with LiNbO$_3$. This observation is inconsistent  with results reported by  Caumes et al \cite{Caumes2002}.  They did not observe  an obvious Kerr-like nonlinearity contributing to $n_2$ due to the phase mismatch  occurs in ZnSe. The reason for our results may be that $\bm{\chi}^{(2)}$ of LiNbO$_3$ is larger than that of ZnSe. Distinguishing Kerr and Kerr-like nonlinearities is important because femtosecond laser pulses interacting with LiNbO$_3$ has important applications such as excitation and imaging of phonon polaritons.

Another aspect should be mentioned  is that we only observed an obvious phonon-polariton signal when the interaction length is small.   The reason is that phonon poaritons can propagate in the crystals.  The phase mismatch also hinders the observation of phonon-polariton signal. When  the interaction length is  small, the phase mismatching effect is  also small. 

It is necessary to compare  the THz related Kerr-like effect and intense THz wave induced Kerr effect. With the great development of intense THz sources, it becomes possible to observe third-order nonlinear effects in the THz frequency range. For example, Zalkovskij et al \cite{Zalkovskij2013}  have observed THz-induced optical birefringence in amorphous chalcogenide glasses which are  centrosymmetric.  It is impossible to observe the THz related Kerr-like effect reported here in amorphous chalcogenide glasses because the $\bm{\chi^{(2)}}$ of the  centrosymmetric material vanishes. Besides, the Kerr-like effect in LiNbO$_3$ is non-instantaneous while the induced change of refractive index in amorphous chalcogenide glasses follows the intensity profile of the THz pulse.

In conclusion, we have conducted optical pump-probe experiments on LiNbO$_3$ with different angles between the pump beam and  the probe beam. Both Kerr and Kerr-like nonlinearities have been observed to contribute to $n_2$ with femtosecond laser pulses interacting with LiNbO$_3$.  Besides, we also have excited and detected the low frequency phonon polaritons.  Our  experimental results agree well with theoretical analysis.

This work is supported by Beijing Natural Science Foundation (4194083), the National Natural Science Foundation of China (61775233, 11827807).

\nocite{*}
\bibliography{aipsamp}% Produces the bibliography via BibTeX.

\end{document}